\documentclass[a4paper,11pt]{article}

\usepackage{graphicx,array}
\usepackage{color}
\usepackage{latexsym}
\usepackage{amsthm}
\usepackage{amsmath}
\usepackage{amssymb}
\usepackage{empheq}

\usepackage{dsfont}
\usepackage{epsfig}
\usepackage{slashed}
\usepackage{bbold}
\usepackage{psfrag}
\usepackage[svgnames]{xcolor}
\PassOptionsToPackage{caption=false}{subfig}
\usepackage{subcaption}
\usepackage{xfrac}
\usepackage{multirow}
\usepackage{booktabs}
\usepackage{cite}
\usepackage[normalem]{ulem}
\usepackage{jheppub} 
\usepackage{adjustbox}

\newcommand{\be}{\begin{equation}}
\newcommand{\ee}{\end{equation}}
\newcommand{\bea}{\begin{eqnarray}}
\newcommand{\eea}{\end{eqnarray}}
\newcommand{\bei}{\begin{itemize}}
\newcommand{\eei}{\end{itemize}}

\title{Illuminating the dark: mono-$\gamma$ signals at NA62}

\author[a]{D. Barducci,}\emailAdd{daniele.barducci@roma1.infn.it}
\author[b,c,d]{E. Bertuzzo,}\emailAdd{enrico.bertuzzo@unimore.it}
\author[e]{M. Taoso,}\emailAdd{marco.taoso@to.infn.it}
\author[f]{C. A. Ternes,}\emailAdd{christoph.ternes@lngs.infn.it}
\author[g]{C. Toni}\emailAdd{claudio.toni.1@unipd.it}

\affiliation[a]{
Dipartimento di Fisica Enrico Fermi, Universit\`a di Pisa and INFN, Sezione di Pisa,
Largo Bruno Pontecorvo 3, I-56127 Pisa, Italy
}
\affiliation[b]{
Dipartimento di Scienze Fisiche, Informatiche e Matematiche, Universit\`a di Modena e Reggio Emilia, via G. Campi 213/A, 41125 Modena, Italy
}
\affiliation[c]{
INFN sezione di Bologna, via
Irnerio 46, 40126 Bologna, Italy
}
\affiliation[d]{
Instituto de F\'isica, Universidade de S\~ao Paulo, C.P. 66.318, 05315-970 S\~ao Paulo, Brazil
}
\affiliation[e]{
Istituto Nazionale di Fisica Nucleare, Sezione di Torino, via P. Giuria 1, I–10125 Torino, Italy
}
\affiliation[f]{
Istituto Nazionale di Fisica Nucleare (INFN),
Laboratori Nazionali del Gran Sasso, 67100 Assergi, L’Aquila (AQ), Italy
}
\affiliation[g]{
Dipartimento di Fisica e Astronomia `G.~Galilei', Universit\`a di Padova and INFN Sezione di Padova, Via F. Marzolo 8, 35131 Padova, Italy
}

\abstract{Dipole interactions between dark sector states or between a Standard Model particle and a dark state can efficiently be searched for via high-intensity fixed-target facilities. 
We propose to look for the associated mono-$\gamma$ signature
at the NA62 experiment running in beam dump mode. Focusing on models of dipole inelastic Dark Matter and active-sterile neutrino dipole interactions, we compute the corresponding expected sensitivities finding promising prospects for discovery already with $\sim10^{17}$ proton-on-target, corresponding to the present accumulated dataset.
}

\begin{document} 
\maketitle

\section{Introduction}

Light new particles, with masses around the GeV scale and 
neutral under the Standard Model (SM) gauge group, are ubiquitous in New Physics (NP) theories featuring a dark sector, and
can be related to the origin of Dark Matter (DM) or neutrino masses, see, {\it e.g.}, Ref.~\cite{Antel:2023hkf} for a recent review. 
Due to their dark nature, they
are typically hard targets for
direct production at collider experiments, but,
on general grounds, can be efficiently tested in
exotic decays of SM mesons, 
since the latter can be copiously produced in high-intensity beam dump experiments. This is the case, {\it e.g.}, of dark photons or
sterile neutrinos that interact with the SM only via mixing, and/or dark Higgses\,\cite{Beacham:2019nyx}. Depending on the decay mode of the NP state, different search strategies can be envisaged, but a generic feature of these signatures is the one of long-lived particles (LLPs) decaying at a macroscopic distance from their production point, a feature 
due to their weakly coupled nature.

While specific light dark sectors 
can be on their own full-fledged ultraviolet theories, at least in the same way as the SM is, one can imagine that, in a more complete DM or neutrino mass theory, additional degrees of freedom might be present at 
scales larger than the electroweak (EW) one. These new states will leave an imprint at low energy that can be parametrized by a set of non-renormalizable operators of dimension $d>4$, built out of SM and light NP fields only.
With these additional effective interactions, more observational possibilities open up to reveal the nature of the dark sectors. 
These deformations have been studied in Ref.\,\cite{Barducci:2021egn} for the dark photon and, {\it e.g.}, in Ref.~\cite{delAguila:2008ir,Liao:2016qyd,Li:2021tsq,Graesser:2007yj,Graesser:2007pc,Caputo:2017pit,Butterworth:2019iff,Barducci:2020icf,Aparici:2009fh,Balaji:2020oig,Barducci:2020ncz,Cho:2021yxk,Delgado:2022fea,Zhou:2021ylt,DeVries:2020jbs,Beltran:2021hpq,Cottin:2021lzz,Barducci:2022gdv,Beltran:2023nli,Beltran:2022ast,Beltran:2023ksw,Gunther:2023vmz,Duarte:2023tdw,Magill:2018jla} for sterile neutrinos.

An interesting and well studied scenario is the one where the new states interact with the photon through dipole operators.  
This is the case of neutrino dipole portal interactions\,\cite{Magill:2018jla}, sterile neutrino dipole models\,\cite{delAguila:2008ir} and dipole inelastic DM (see, {\it e.g.}, Refs.~\cite{Dienes:2023uve,Masso:2009mu,Chang:2010en,Izaguirre:2015zva}).

The dipole interaction can trigger both the production of a light new particle and its decay.
Production occurs for instance via meson decay and possibly upscattering of an active neutrino into a sterile one,
while the decay of the new state can produce a final state with a mono-$\gamma$ signature. Various works have shown how to exploit this possibility in order to test these dark sector theories
at different proposed future experiments, such as FASER, FASER$\nu$, SHiP or DUNE\,\cite{Chu:2020ysb,Barducci:2022gdv,Barducci:2023hzo,Dienes:2023uve,Magill:2018jla,Jodlowski:2020vhr,Ovchynnikov:2022rqj,Schwetz:2020xra,Atkinson:2021rnp}.

There exists, however, an additional experimental facility which, albeit originally designed for a different purpose, has already collected and still will collect more data in beam dump mode with the goal of testing dark sector theories. 
This is the case of the NA62 experiment which, although 
originally built to search for the SM rare decay $K^+ \to \pi^+ \nu \bar \nu$\,\cite{NA62:2017rwk}, 
can be also exploited for more generic searches of NP once it will turn to the programmed beam dump mode, see {\it e.g.}~\cite{NA62:2023qyn} for a recent analysis.
The goal of our work is to consider all the  
dipole interactions previously mentioned and 
show how they can be efficiently tested in the NA62 experiment by a dedicated analysis searching for a mono-$\gamma$ signal in its decay volume placed at around 79~m from the interaction point. For comparison, we will also show prospects for the detection of the mono-$\gamma$ signal at the future beam dump experiment SHiP\,\cite{SHiP:2015vad,SHiP:2021nfo} and at the proposed experiment FASER 2\,\cite{Feng:2017uoz,Feng:2022inv}  at the Large Hadron Collider (LHC), as well as various bounds that limit the parameter space of the effective operators considered. 

The paper is organized as follows. In Sec.~\ref{sec:model} we introduce the interactions analyzed in this study while in Sec.~\ref{sec:exp} we describe the experimental facilities under consideration. Our results are presented in Sec.~\ref{sec:res} and we conclude in Sec.~\ref{sec:conc}. We also add an Appendix~\ref{app:decay_widths} where we collect details on the calculations of the decay widths used for our analysis.

\section{Benchmark models}\label{sec:model}

Our phenomenological analysis will focus on the 
following two different scenarios, which share the same underlying structure of a dipole portal interaction. The relevant Dirac structure is $\sigma^{\mu\nu} = i[\gamma^\mu, \gamma^\nu]/2$ which, being antisymmetric in the spinor indices, can only involve two different Majorana spinor fields (or, when written using two-component spinors, the dipole interaction must involve two different Weyl fermions). The scenarios we will consider are:
\begin{itemize}
    \item The dark-dark dipole 
    \be\label{eq:dipole_sterile-sterile}
        \mathcal{O}_\chi^5 = d\, \bar{\chi}_2 \sigma^{\mu\nu} \chi_1 \, F_{\mu\nu},
    \ee
    where $\chi_{1,2}$ are two Majorana NP states, completely neutral under the SM gauge group, $F_{\mu\nu}$ is the photon field-strength tensor and $d$ is the Wilson coefficient of the operator. These states can represent an inelastic DM pair\,\cite{Dienes:2023uve} or a pair of sterile neutrinos\,\cite{Aparici:2009fh}. The phenomenology  at laboratory experiments of the two cases is the same in the limit in which the mixing between active and sterile neutrinos is taken to be negligible, as we will always assume to be the case in this paper, see~\cite{Barducci:2022gdv} for a discussion on this point. We take $\chi_1$ to have mass $M_1$ and $\chi_2$ to have mass $M_2$, with $M_1 < M_2$. We quantify their relative splitting with the parameter
    \be
        \delta = \frac{M_2 - M_1}{M_1}.
    \ee
    \item The active-sterile neutrino dipole
    \be\label{eq:dipole_active-sterile}
        \mathcal{O}^5_{\nu,i} = d_{i}\, \bar \nu_{i,L} \sigma^{\mu\nu} N_{1,R} F_{\mu\nu} ,
    \ee
    between an active neutrino of flavor $i=e,\mu,\tau$ and a single sterile neutrino $N_1$, taken to be a Dirac state\,\cite{Magill:2018jla}. Again, $d_{i}$ is the Wilson coefficient of the operator.
\end{itemize}

For an example of how to generate the dipole operator involving two dark states, see\,\cite{Aparici:2009oua}. For considerations on how to UV complete the $\mathcal{O}^5_{\nu,i}$ operator, see\,\cite{Magill:2018jla}. We observe that these UV completions will generate dipole interactions with the hypercharge field-strength  tensor, which in the broken EW phase will contain both the photon and the $Z$-boson field strengths. Since we are interested in phenomena that happen at energies well below the EW scale, we will consider only the phenomenology involving photons, ignoring the dipole operators with a $Z$-boson. Moreover, it should be noted that while the operator $\mathcal{O}^5_{\chi}$ is generated in UV complete models directly at $d=5$, the $\mathcal{O}^5_{\nu,i}$ operator is typically generated at $d=6$, in the gauge invariant form $\bar{L} \tilde{H}\sigma^{\mu\nu} N_{1,R} B_{\mu\nu}$, where $L$ and $\tilde{H} = i \sigma_2 H^*$ are the lepton and conjugate Higgs doublets, respectively, while $B_{\mu\nu}$ is the hypercharge field-strength tensor. Finally, we observe that dipole operators are necessarily produced at the one (or higher) loop level in any explicit weakly coupled UV completion, so that explicit loop factors may be needed to correctly estimate the size of the dipole coefficients. To facilitate the comparison with the existing literature, however, in the remainder of the paper we will completely encode the effect of the dipoles in a coefficient $d$, regardless of their UV origin.
\begin{figure}
	\centering
	\includegraphics[scale=0.6]{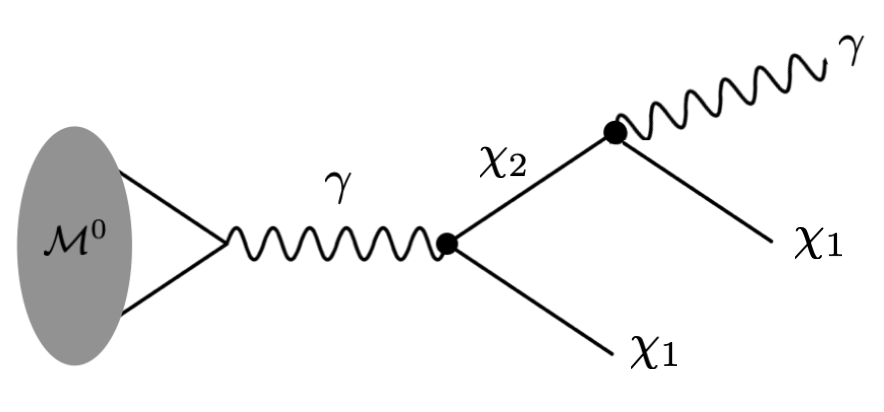}
	\caption{A mono-$\gamma$ final state produced by the decay of a SM meson ${\cal M}^0$ through the operator $O^5_\chi$ of Eq.~\eqref{eq:dipole_sterile-sterile}. A similar diagram is obtained in the case of the ${\cal O}^5_{\nu,i}$ operator,
    with a trivial replacement of the involved particles.}
	\label{fig:diagram}
\end{figure}

In this work, we focus on the possibility of testing the interactions of Eq.\,\eqref{eq:dipole_sterile-sterile} and Eq.\,\eqref{eq:dipole_active-sterile} taking advantage of the fact that, after the $\chi_1\chi_2$ (or $\nu N_1$) pair is produced, the heavier state can travel a macroscopic distance before decaying inside the detector. More in detail, we exploit the dipole interactions both to produce the $\chi_1\chi_2$ ($\nu N_1$) pair via meson decays, and to generate the mono-$\gamma$ signal $\chi_2 \to \chi_1 \gamma$ ($N_1 \to \nu \gamma$). A sketch of the relevant phenomenological process is shown in Fig.\,\ref{fig:diagram} for ${\cal O}^5_\chi$. 

\section{NA62 and other experiments under consideration}\label{sec:exp}

The main purpose of our work is to analyse what region of parameter space of the interactions of Eq.\,\eqref{eq:dipole_sterile-sterile} and Eq.\,\eqref{eq:dipole_active-sterile} can be probed by the NA62 experiment.
This experiment has been designed as a $K$ factory to look for the decay $K^+ \to \pi^+ \nu \bar \nu$, a flavor-changing neutral current (FCNC) that, at lowest order in the SM, proceeds via electroweak box and penguin diagrams, both dominated by $t$-quark exchange.
Being both GIM and CKM suppressed, this process has an extremely small branching ratio, whose SM prediction is ${\rm BR}=(8.4\pm 1.0)\times 10^{-11}$~\cite{Buras:2015qea}. In its normal operation mode, NA62 exploits a 400\,GeV proton beam extracted from the CERN SPS imprinting on a $40\,$cm  beryllium rod. The resulting secondary hadron beam of positively charged particles consists of $\pi^+$ ($\sim 70\%$), protons ($\sim 23\%$) and $K^+$ ($\sim 7\%$), with a nominal energy of about 75\;GeV. The  positively charged $K$ beam is then selected and decays in a vacuum chamber, with dedicated detectors that ensure an hermetic photon coverage to suppress the large $\pi^0\to \gamma\gamma$ rate from $K^+ \to \pi^+ \pi^0$ and ensure sensitivity to the $K^+ \to \pi^+ \nu \bar \nu$ decay. This decay mode has recently been observed for the first time with a rate in agreement with the SM expectation\,\cite{NA62:2021zjw}.
Since the process $K^+ \to \pi^+ \nu \bar \nu$ is an extremely rare decay, NA62 is an ideal place also to look for NP effects that can produce a similar final state. For example,  searches for $\pi^0 \to {\rm invisible}$ have been performed~\cite{NA62:2020pwi}. The large flux of $K$ mesons also allows to look for complementary signatures, such as 
those of additional light states produced in $K^+$ decays.
This is the case of axion-like particles\,\cite{NA62:2023olg} and 
heavy neutral leptons mixing with the SM neutrinos\,\cite{NA62:2020mcv,NA62:2021bji}. Common to all these analyses is the high photon identification and rejection 
rate that can be achieved by NA62 which, as mentioned, is crucial to veto on the huge diphoton rate arising from $\pi^0$ decays and that could be exploited to target the signature of Fig.\,\ref{fig:diagram}.  

Beyond the 
default Kaon mode, the NA62 experiment has the capabilities of running in beam dump mode, with little modifications 
to its 
experimental design. Running in  beam dump mode opens the possibility of testing light and weakly coupled states by directly producing them 
from the decay of mesons copiously produced at NA62 or through other processes,
extending the reach beyond the $K$ mass and exploiting the decay of multiple SM mesons.
The first NP results in this operation mode have recently appeared with a dataset of around $10^{17}$ protons on target (POT)~\cite{Jerhot:2023dhn,NA62:2023nhs}, setting bounds on dark photons and axion-like particles. 
Future runs plan to collect a larger dataset of ${\cal O}(10^{18})$ POT, which can be further augmented to ${\cal O}(5\times 10^{19})$ if the future high-intensity Kaon experiment (HIKE) will be realized~\cite{HIKE:2022qra}\,\footnote{While at the time of writing the future of HIKE is uncertain, in the following we will show prospects also for this design, in order to quantify the sensitivity that can be achieved with a larger dataset.}. 

We now summarize the NA62 experimental specifications and geometry that will be used throughout our analysis. For comparison, we will also compute the projected reach of other future proposed experiment on the mono-$\gamma$ final state, such as SHiP, HIKE and FASER 2. We briefly describe also their experimental design in this section. We highlight again that the signal we are investigating is the one of a SM meson produced in the interaction point (IP), promptly decaying into dark sector particles which are long lived and that decay displaced from the IP into a final state consisting of a single photon and missing energy.

\paragraph{NA62 in beam dump mode:}

In the beam dump mode of NA62, a 400\;GeV proton beam from the CERN SPS is dumped on a copper target producing a flux of SM mesons, see~\cite{NA62:2023qyn}. 
The NA62 decay volume starts at $\sim$79\,m from the IP. It has a cylindrical shape with a radius of 1\,m and it is aligned with the beam axis. 
After 138\,m from the beginning of the decay volume there is a cylindrical electromagnetic calorimeter (eCAL) spanning around 1.2 m in radius.
For our analysis we require the exotic particles
to decay producing a single photon signature between $L_i = 79\,$m and $L_f=(79+138)\,{\rm m}=217\,$m from the IP. The maximum allowed polar angle for the LLP is required to be $\theta_{\rm max}=\arctan\frac{1.2}{217}\simeq 5.5 \times 10^{-3}$, in order for its trajectory to intersect the eCAL\,\footnote{Strictly speaking, it is the photon trajectory that should hit the eCAL. However, due to the large boost of the dark sector particles, the LLPs result almost collinear with the final photons so that our approximation holds with good accuracy.
}. 
Regarding the integrated dataset, we will consider a total number of POT $N_{\rm POT}=10^{18}$.
The NA62 experiment will run until the LHC long shutdown 3, which is planned around the year 2026.
We will also consider a smaller dataset, namely $N_{\rm POT}=10^{17}$, in order to derive the expected sensitivity with the dataset already collected in beam dump mode or within a shorter timescale for future data taking.

\paragraph{HIKE:}
After the long shutdown 3 there is the proposal of hosting a high-intensity Kaon experiment (HIKE) within the same infrastructure now occupied by NA62\,\cite{HIKE:2022qra,HIKE:2023ext,Hike3}. The feasibility of this option is uncertain and currently under discussion. In computing the projected reach of the HIKE experiment, we use the same geometry as for the NA62 case, see also~\cite{Ovchynnikov:2023cry}, and increasing the statistics to $N_{\rm POT}= 5\times 10^{19}$. Whether or not the HIKE proposal will be realized, we find it instructive to see what could be achieved with larger statistics with the NA62 detector design. 

\paragraph{SHiP:} This is another proposal for a future beam dump experiment to be hosted at CERN.
To implement its geometry, 
we follow the recent ECN3 proposal\,\cite{SHiPNeWnote}, updating the results that some of us obtained in a previous paper\,\cite{Barducci:2022gdv}. The SHiP experiment also uses a 400\,GeV proton beam, dumped on a molybdenum/tungsten target. 
The distance of the decay volume from the IP is taken to be $L_i=33\,{\rm m}$, while the decay volume has a length of 50\;m, $L_f=85\,{\rm m}.$ The SHiP decay volume has a squared-base pyramid shape, with upstream and downstream areas of $1.0\times 2.7\,$m$^2$ and $4.0\times 6.0\,$m$^2$, respectively.  
For simplicity we approximate the SHiP geometry to be a cone shape, and we consider a maximum acceptance angle from the beam axis of $\theta_{\rm max} = 2.8\times 10^{-2}$. 
Following\,\cite{SHiPNeWnote}, we consider an exposure of $N_{\rm POT}= 6\times 10^{20}$.

\paragraph{FASER 2:} This is a proposed experiment that should upgrade FASER\,\cite{Feng:2017uoz,FASER:2018eoc}, currently installed along the beam axis near the ATLAS experiment. The geometry of FASER 2 is as follows\,\cite{Feng:2022inv}: the detector will be placed at a distance of 620 m from the IP. The decay volume consists of a cylinder with a radius of 1\,m and a length of 10\,m, followed by a tracking system of about 10 m and then a calorimeter.
In our analysis we thus require the heavier dark state to decay between $L_i=$ 620\,m and $L_f = (620+10+10)$\,m = 640\,m. We thus require the maximum polar angle between the direction of the LLP and the beam axis to be $\theta_\text{max} = \text{arctan}\frac{1}{640} = 1.6\times 10^{-3}$. FASER 2 is expected to collect a total luminosity of 3\,ab$^{-1}$, the expected integrated luminosity of the high-luminosity LHC.

\section{Projected sensitivities and constraints}\label{sec:res}

In this section we discuss the computation of the NA62 sensitivity to the dipole interactions discussed in Sec.\,\ref{sec:model} and the comparison with the other experiments mentioned above. We start with an explanation of how we compute the sensitivity; we then discuss current limits on the parameter space. Finally, we present our results, summarized in Figs.\,\ref{fig:sensInelastic}-\ref{fig:sensActiveSterile}.

\subsection{Sensitivity estimation}

In order to estimate the sensitivity reach 
of the experiments described above
to the Wilson coefficients of the effective operators of Eq.\,\eqref{eq:dipole_sterile-sterile} and Eq.\eqref{eq:dipole_active-sterile}, we proceed in the following way. 
We compute the total number of signal events according to
\be\label{eq:Nevents}
N_{\rm evts} = \sum_{{\cal M}} N_{{\cal M}}\times {\rm BR}_\mathcal{M}\times f_{\rm dec}\times \epsilon_{\rm sel} \ ,
\ee
where 
\begin{itemize}
    \item $N_\mathcal{M}$ is the total number of mesons $\mathcal{M}$ produced at the experiment considered, given by
\be
N_\mathcal{M} = \left\{
\begin{array}{lcl}
N_{\rm POT} \times f_\mathcal{M}, & ~~~~~& \text{for NA62,\,HIKE,\,SHiP,}\\
\sigma_\text{inel}\times \mathcal{L} \times f_\mathcal{M} & ~~~~~& \text{for FASER\,2,}\\
\end{array}\right.
\ee
where $N_{\rm POT}$ is the total number of POT collected by the fixed-target experiment, $\sigma_\text{inel} = 79.5$~mb is the inelastic proton-proton cross section at the LHC with $\sqrt s = 13$~TeV\,\cite{TOTEM:2017asr}, $\mathcal{L} = 3$~ab$^{-1}$ is the corresponding total integrated luminosity at the end of the high-luminosity phase, and $f_\mathcal{M}$ is the average number of mesons (or multiplicity) produced in each proton-target interaction (proton-proton for FASER\,2). 

For FASER\,2, the momentum and energy distributions of the different mesons, as well as their multiplicities, are taken from the {\tt FORESEE} package~\cite{Kling:2021fwx}.
Instead, for the fixed-target experiments, we simulate the production of the different mesons in proton-proton collisions with {\tt PYTHIA8}\,\cite{Sjostrand:2007gs,Bierlich:2022pfr}.
Then, to take into account that the considered experiments collide protons over the nuclei of the target, we rescale the multiplicities of the charmed and $B$ mesons by a factor $A^{0.29}$, with $A$ the nuclear mass number. This scaling originates from the dependence $\sim A^{0.71}$ of the inelastic proton-target cross section~\cite{Carvalho:2003pza,SHiP:2015vad} and an approximate linear dependence on $A$ of the charm and beauty production cross-sections~\cite{Lourenco:2006vw,SHiPnote}.
For the charged mesons $\mathcal{M}^\pm = \left\{B^\pm, D^\pm, D_s\right\}$, the multiplicities are taken from~\cite{SHiP:2018xqw}.
We collect the multiplicities for $p-p$ scattering that we used in our computation in Tab.\,\ref{tab:multiplicities}. As mentioned above, for charmed and bottomed mesons it is necessary to multiply these numbers by $A^{0.29}$ to obtain the total number of mesons produced at NA62/HIKE and SHiP. 
When available, we have compared the multiplicities obtained with this procedure to those quoted in the {\tt SensCalc} package\,\cite{Ovchynnikov:2023cry}, finding reasonable agreement.
Charged kaons are abundantly produced in proton beam dump experiments but most of them are absorbed in the target material before decaying. Ref.~\cite{Gorbunov:2020rjx}, by means of a GEANT4 simulation, found that (at SHiP) in average each proton interaction leads to 0.29 $K^+$ and 0.07 $K^-$  which decay in flight, {\it i.e.} that they are not absorbed before decaying, and are not strongly attenuated.
We include $K^{\pm}$ in our simulation using this normalization (neglecting small variations which will be present for NA62 due to the different detector material), using the angular and momentum distributions of the $K^{\pm}$ mesons obtained from a {\tt PYTHIA8} simulation.
We have found that charged kaons give to the event rate a contribution comparable to the one of other mesons we considered,  when their decay into dark particles is kinematically open. On the other hand, the sensitivities in Fig.~\ref{fig:sensActiveSterile} are only slightly reduced if $K^{\pm}$ are neglected because of the strong dependence of the number of signal events on $\Lambda$.

\item BR$_\mathcal{M}$ is the branching ratio for the decay of the meson $\mathcal{M}$ into dark particles, obtained using the widths collected in Appendix\,\ref{app:decay_widths}.
\item $f_{\rm dec}$ is the fraction of dark states decaying inside the decay volume of the experiment, or more in general in a region of the detector where the photon signal can in principle be detected. It is given by
\be
f_{\rm dec} = e^{-\frac{L_{i}}{\beta\gamma c \tau}} - e^{-\frac{L_{f}}{\beta\gamma c \tau}} \ ,
\ee
where $\beta \gamma c \tau$ is the decay length of the heavier dark state in the laboratory frame, and $L_{i,f}$ are the minimum and maximum distances from the IP at which the heavier dark state is required to decay to produce a detectable photon signal, see Sec.\,\ref{sec:exp}.
The decay widths for the processes $\chi_2 \to \chi_1 \gamma$ or $N_1 \to \nu_i \gamma$ are given in Appendix\,\ref{app:decay_widths}.
\item Finally, $\epsilon_{\rm sel}$ is the selection acceptance for the photon signal,
which depends on the detector geometry as well as on the 
threshold imposed on the photon energy.
We estimate this quantity with a Monte Carlo simulation,
generating the spectra of the relevant mesons via {\tt PYTHIA8}~\cite{Sjostrand:2007gs,Bierlich:2022pfr} and the {\tt FORESEE} package~\cite{Kling:2021fwx}, and implementing the detector geometries described in Sec.\,\ref{sec:exp}.
\end{itemize}
We observe that, for the operator $\mathcal{O}^5_\chi$, 
the dark states are dominantly produced by the decay of neutral vector  mesons mediated by an off-shell photon, $\mathcal{M}^0 \to \chi_1 \chi_2,$ and by the decay of the light pseudoscalar neutral mesons $\mathcal{M}_P^0 \to \chi_1 \chi_2\gamma$ via the chiral anomaly and a subsequent photon splitting into dark states.
In our analysis we consider $\mathcal{M}^0= \left\{\rho, \omega, J/\psi, \Upsilon, \phi\right\}$ and $\mathcal{M}_P^0= \left\{\pi^0, \eta, \eta^{\prime}\right\}$.
For the operator $\mathcal{O}^5_{\nu_i}$, 
an additional channel is provided by the decay of
the charged mesons $\mathcal{M}^\pm \to e^\pm N_1 \gamma$ via an off-shell active neutrino. We include the charged mesons $\mathcal{M}^\pm = \left\{B^\pm, D^\pm, D_s, K^{\pm}\right\}$.

We conclude this section with an important observation. When we consider the active-sterile dipole operator of Eq.\,\eqref{eq:dipole_active-sterile}, we assume that $N_1$ is a Dirac fermion\,\cite{Magill:2018jla}. This means that, in addition to the production of $N_1$ by meson decays and its subsequent decay, in the computation of the total number of events using Eq.\,\eqref{eq:Nevents} we must also consider the production and decay of its antiparticle $\bar{N}_1$.

\begin{table}
    \centering
    \begin{adjustbox}{width=1\textwidth}
    \begin{tabular}{ | c | c | c | c | c | c | c | c | c | c | c |}
    \hline
         $f_{\pi^0}$ & $f_{\eta}$ & $f_{\eta^{\prime}}$ & $f_\rho$ & $f_\omega$ & $f_{\phi}$ & $f_{J/\Psi}$ & $f_{\Upsilon}$ & $f_{D^\pm}$ & $f_{D_s}$& $f_{B^\pm}$  \\
        \hline
         4.3 & 0.49 & 0.055 & 0.58 & 0.57 & 0.021 & $4.7\times 10^{-6}$ & $2.2\times 10^{-9}$ & $4.3\times10^{-4}$ & $1.8\times10^{-4}$ & $6.0\times10^{-8}$ \\
         \hline
    \end{tabular}
    \end{adjustbox}
    \caption{Expected multiplicities for meson production at proton-proton interactions relevant for the experiments NA62/HIKE and SHiP.
   The multiplicity of kaons is reported and commented in the text.
    }
    \label{tab:multiplicities}
\end{table}

\subsection{Current limits}
\label{sec:Currentlimits}
Before presenting the sensitivity of the experiments discussed above
on 
the operators in Eq.\,\eqref{eq:dipole_sterile-sterile} and Eq.\,\eqref{eq:dipole_active-sterile}, we describe 
existing limits on their parameter space. 

The active-sterile neutrino dipole operator in Eq.\,\eqref{eq:dipole_active-sterile} is constrained by a variety of laboratory searches as well as astrophysical and cosmological observations. 
We report and show the corresponding strongest limits in Fig.\,\ref{fig:sensActiveSterile}. More specifically, we show bounds from LEP~\cite{Magill:2018jla}, LSND~\cite{Magill:2018jla}, NOMAD~\cite{Magill:2018jla}, MiniBooNE~\cite{Magill:2018jla}, Super-Kamiokande atmospheric neutrino data (SK)~\cite{Gustafson:2022rsz}, Borexino and Super-Kamiokande (Borexino+SK)~\cite{Plestid:2020vqf}, DONUT~\cite{DONUT:2001zvi}, CHARM-II~\cite{Coloma:2017ppo}, BBN~\cite{Brdar:2020quo} and Supernova SN1987A~\cite{Chauhan:2024nfa} (assuming lepton flavor universality).
It is important to observe that most of these constraints exploit intense neutrino beams in laboratory experiments or astrophysical neutrino sources, which can then lead to the production of sterile neutrinos through the operator of Eq.\,\eqref{eq:dipole_active-sterile}.
Therefore these limits do not directly apply to the dark-dark dipole operator in Eq.\,\eqref{eq:dipole_sterile-sterile}, for which we present below a discussion of the relevant constraints from beam dump (BEBC, CHARM-II and NuCal) and collider (BaBar) experiments.
Additional bounds are discussed in~\cite{Barducci:2022gdv}.

Before turning to this discussion, it is important to comment on the possible cosmological relevance of the lightest dark-state, which is completely stable within the simplified scenario of Eq.\,\eqref{eq:dipole_sterile-sterile}. This may not be the case when $\chi_{1,2}$ are identified with two sterile neutrinos, since in this case the fate of the lightest particle ultimately depends on its mixing with the active sector. In particular, the situation where $\chi_1$ decays at an epoch close to the big bang nucleosynthesis epoch is particularly dangerous, but relevant bounds can be evaded, without affecting the phenomenology discussed here, at the price of a certain amount of fine-tuning\,\cite{Barducci:2022gdv}. On the other hand, if the lightest state is completely stable, it can play the role of a DM candidate.
In presenting our results we therefore also show the regions of parameter space where the dark-dipole model can reproduce the cosmological DM abundance measured by Planck\,\cite{Planck:2018vyg} through the thermal freeze-out mechanism. These results are obtained by following the procedure outlined in\,\cite{Griest:1990kh} and cross-checked with the results of\,\cite{Dienes:2023uve}, which are obtained by solving the full set of coupled Boltzmann equation and to which we refer for a detailed discussion.
Let us notice that thermal DM is strongly disfavored for masses below $\lesssim 10$ MeV by CMB and BBN constraints~\cite{Sabti:2019mhn}.
Instead, bounds from direct and indirect DM searches are rather weak in the parameter space of our interest, see~\cite{Dienes:2023uve} for a detailed discussion.

\paragraph{Past beam dump experiments: CHARM-II, BEBC and NuCal} 
We derive constraints on the operator in Eq.\,\eqref{eq:dipole_sterile-sterile} from  CHARM-II, see also~\cite{Chu:2020ysb}. This experiment exploited a 450\,GeV proton beam dumped on a beryllium target and collected $N_{\rm POT}=4.5\times 10^{19}$. The calorimeter detector was placed at 870\,m from the IP, with a transverse area of $3.7\times3.7\,{\rm m}^2,$ and a length of $35.6\,{\rm m}$~\cite{CHARM-II:1989nic}. We consider the analysis of~\cite{CHARM-II:1994dzw}, which was based on a search of a single forward scattered electron producing an electromagnetic shower in the calorimeter, and required electron energies $E\in[3,24]$ GeV. 
We derive 95\% confidence level (CL) bounds by considering a total number of 5429 observed events from neutrino scattering, where the upper limit on the number of signal events has been computed following Appendix B of~\cite{Magill:2018jla}.

We recast a search of single forward scattered electron at the BEBC experiment~\cite{Cooper-Sarkar:1991vsl}, again assuming that this signature can be mimicked by the mono-photon signal.
A 400\,GeV proton beam has been dumped on a copper target, accumulating $N_{\rm POT}=2.72\times10^{18}.$
The detector was located at 404\,m from the IP, with a length of $1.85$\,m and a transverse area of $3.57\times2.52$\,m$^2.$
To mimic the selection criteria of~\cite{Cooper-Sarkar:1991vsl}, we impose a minimum photon energy of 1\,GeV. We derive a 95\%\,CL upper limit from the single electron event reported in~\cite{Cooper-Sarkar:1991vsl}, with an expected background of $0.5\pm 0.1$ events.

The NuCal experiment is based on a 70 GeV proton beam dumped on an iron target. The cylindrical decay volume of a diameter of 2.6\,m and a length of 23\,m was placed 64\,m from the target. 
We follow the analysis of~\cite{Blumlein:2011mv}, based on a search with an accumulated number of protons on target $N_{\rm POT}=1.71\times 10^{18}$. A minimum photon energy of $3\,{\rm GeV}$ is imposed and the reconstruction efficiency is 70\%. A 95\%\,CL upper limit is derived  considering that 5 events were observed, with an estimated background of 3.5 events from neutrino interactions.

\paragraph{BaBar} Data collected in $e^+e^-$ collisions at a center of mass energy around the $\Upsilon$ mass also allow to put bounds on the parameter space of the dark-dark operator. More specifically, we consider the process 
$e^+ e^-  \to \gamma \chi_1 \chi_2$ followed by $\chi_2 \to \gamma \chi_1$ 
and perform a simplified recast of the 
mono-$\gamma$ + missing energy
search of Ref.\,\cite{BaBar:2008aby}, see also\,\cite{Izaguirre:2015zva,Dienes:2023uve}. For this search, BaBar collected 60 fb$^{-1}$ of data. 
We impose the following cuts: $E_\gamma \geq 2$\,GeV for the leading photon, $E_\gamma \geq 20$\,MeV for the sub-leading photon and missing energy $\slashed{E} \geq 50$\,MeV. The cut on the leading photon energy is needed for the BaBar mono-$\gamma$ trigger, while the remaining cuts should allow to bring the background to negligible levels.
Following~\cite{Izaguirre:2015zva}, the sensitivity of this search is estimated requiring 10 signal events.

\begin{figure}
	\centering
	\includegraphics[scale=0.46]{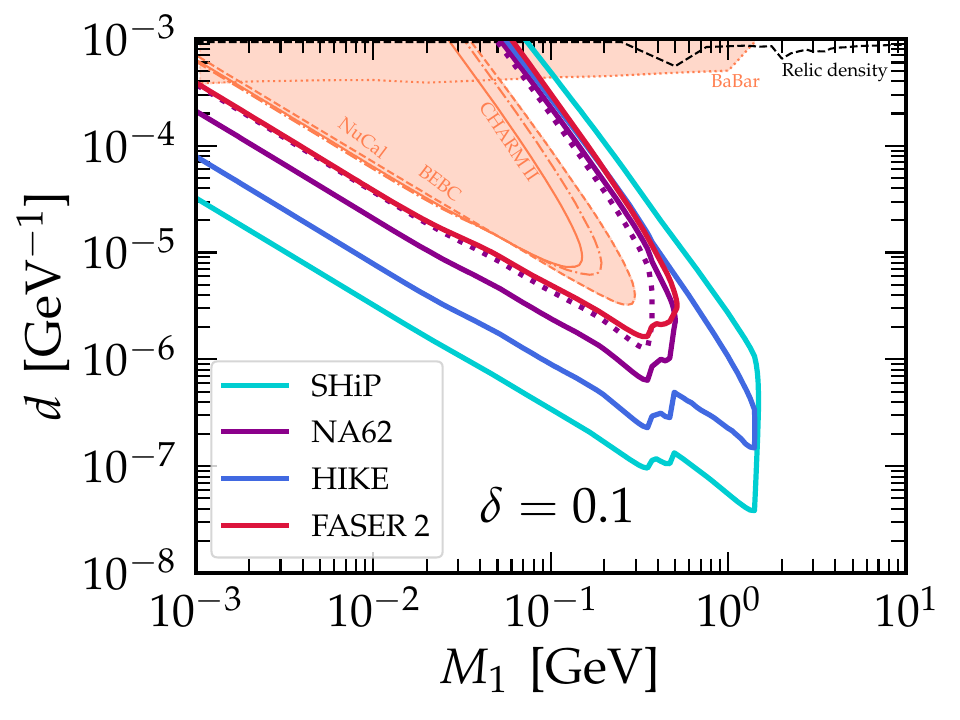}
    \includegraphics[scale=0.46]{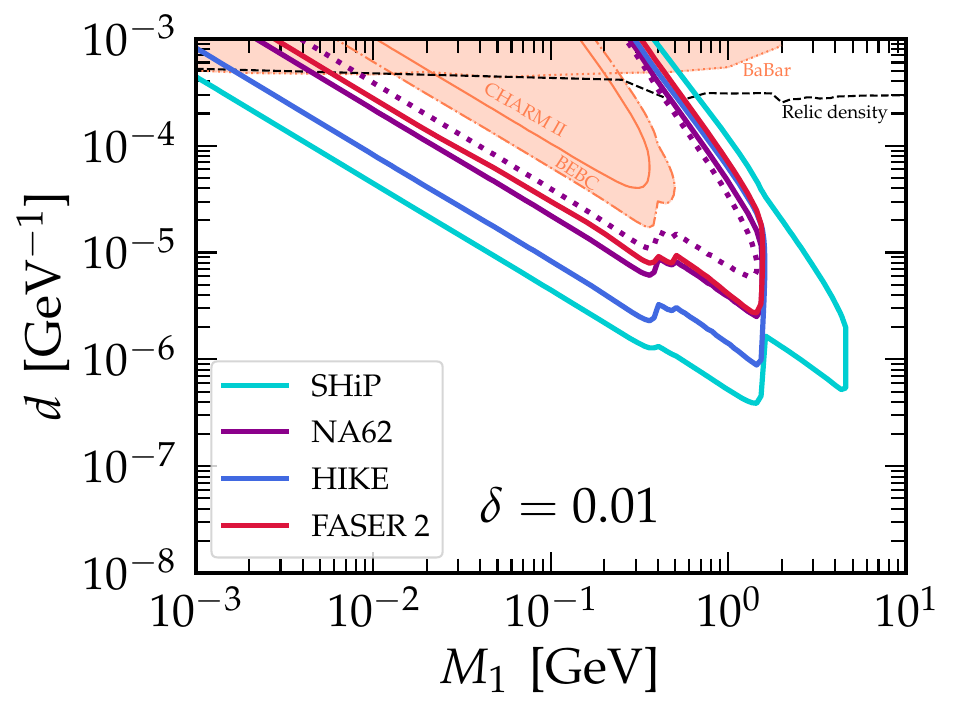}\\
    \includegraphics[scale=0.46]{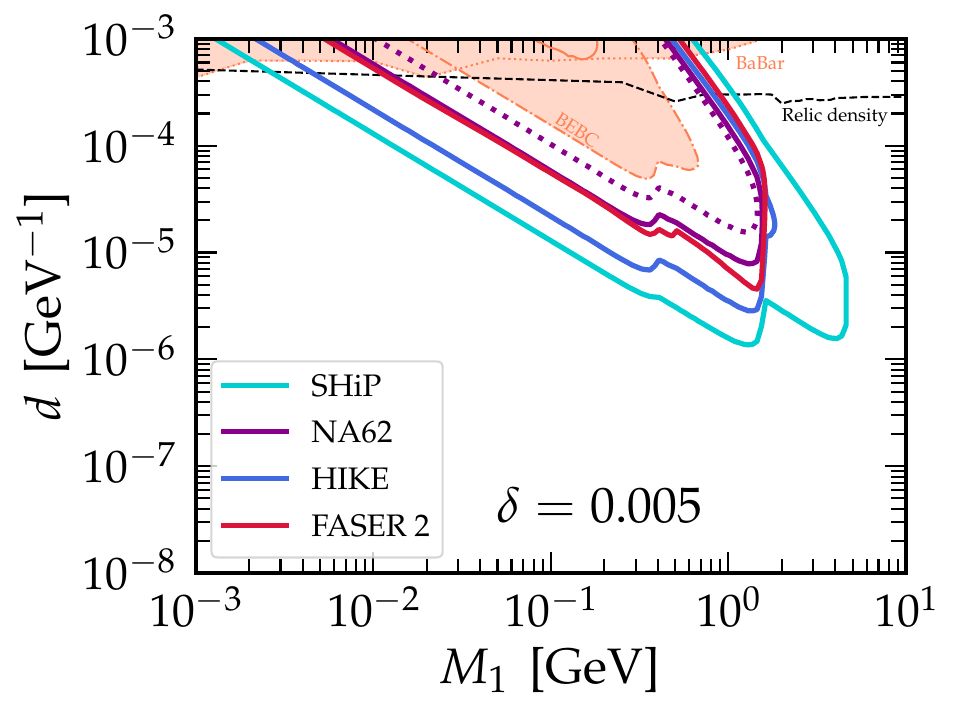}
	\caption{Isocontours for $N_{\rm evts}=3$ signal events induced by the dark-dark dipole operator of Eq.\,\eqref{eq:dipole_sterile-sterile} expected at NA62, SHiP, FASER 2 and HIKE. 
For NA62 we present the sensitivities for $N_{\rm POT}=10^{18}$ (expected completed dataset, solid lines) and $N_{\rm POT}=10^{17}$ (approximately corresponding to the currently collected dataset, dotted lines).
 The colored regions correspond to constraints from Babar (dotted), CHARM II (solid), BEBC (dot-dashed) and NuCal (dashed), see Sec.~\ref{sec:Currentlimits} for more details.
Dashed black lines show the region of the parameter space where the thermal relic density of the lightest dark state matches the DM cosmological abundance.
 }
	\label{fig:sensInelastic}
\end{figure}

\begin{figure}
	\centering
	\includegraphics[scale=0.45]{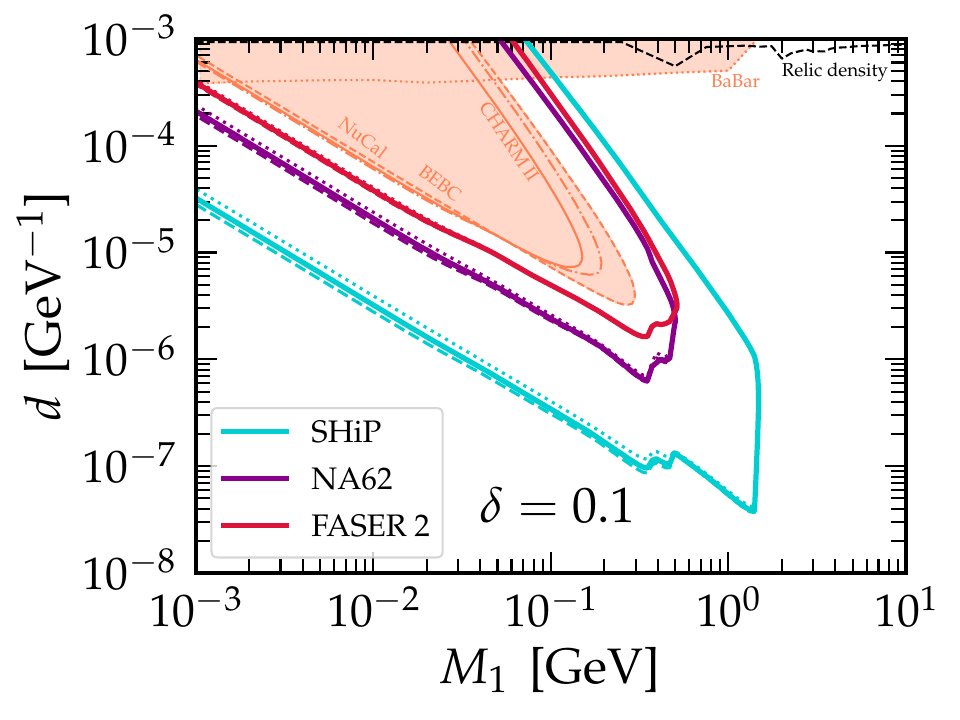}
    \includegraphics[scale=0.45]{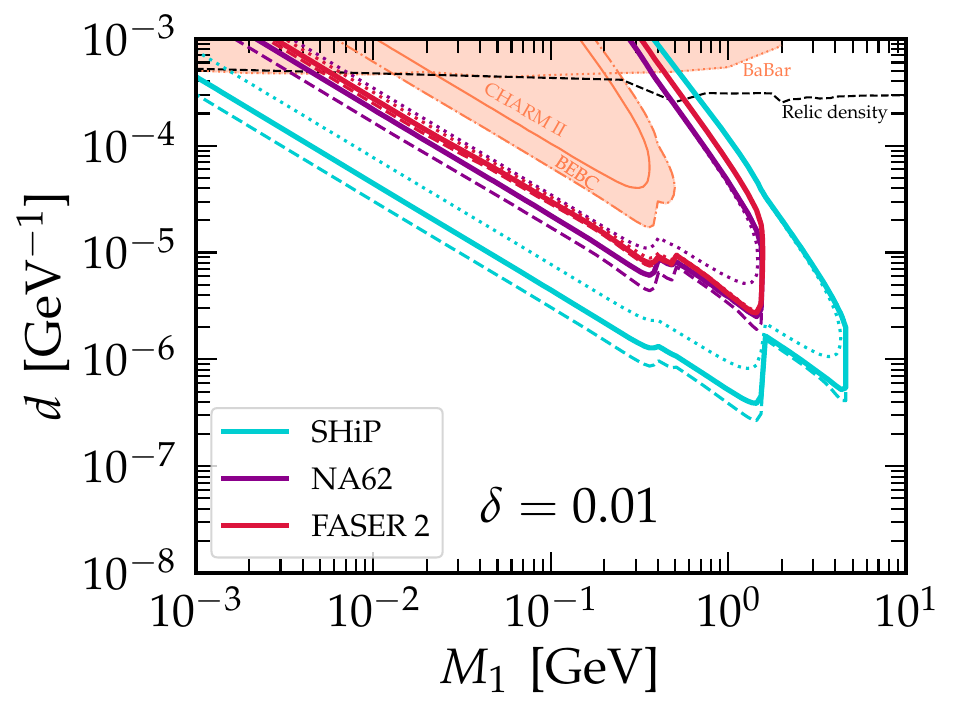}\\
    \includegraphics[scale=0.45]{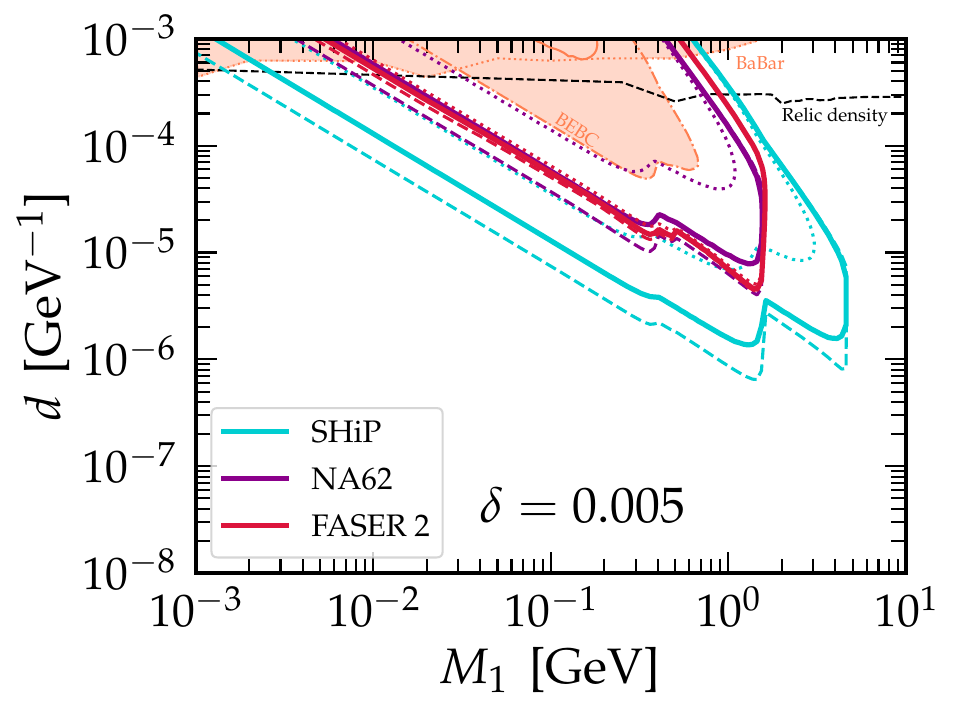}
	\caption{Same as Fig.~\ref{fig:sensInelastic} but considering a threshold for the energy of the photon $E_{\gamma}\ge\,0.5,\,1,\,2\,\rm{GeV}$ respectively for the dashed, solid and dotted lines. To avoid clutter in the figure, we do not show the case of the HIKE experiment here.}
	\label{fig:sensInelasticEcut}
\end{figure}

\begin{figure}[ht]
	\centering
	\includegraphics[scale=0.45]{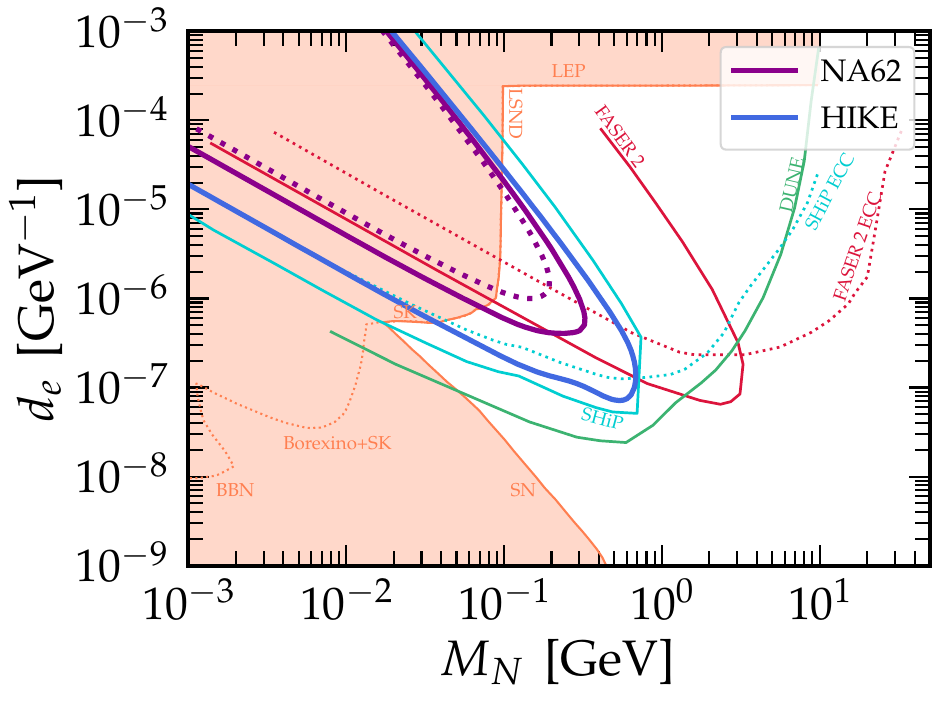}
    \includegraphics[scale=0.45]{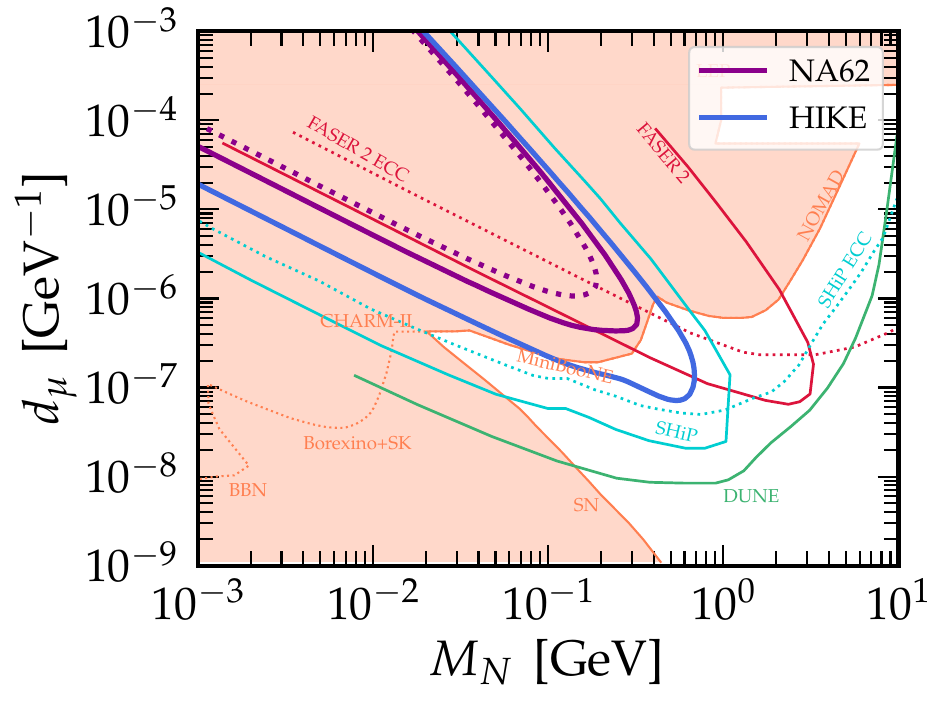}\\
    \includegraphics[scale=0.45]{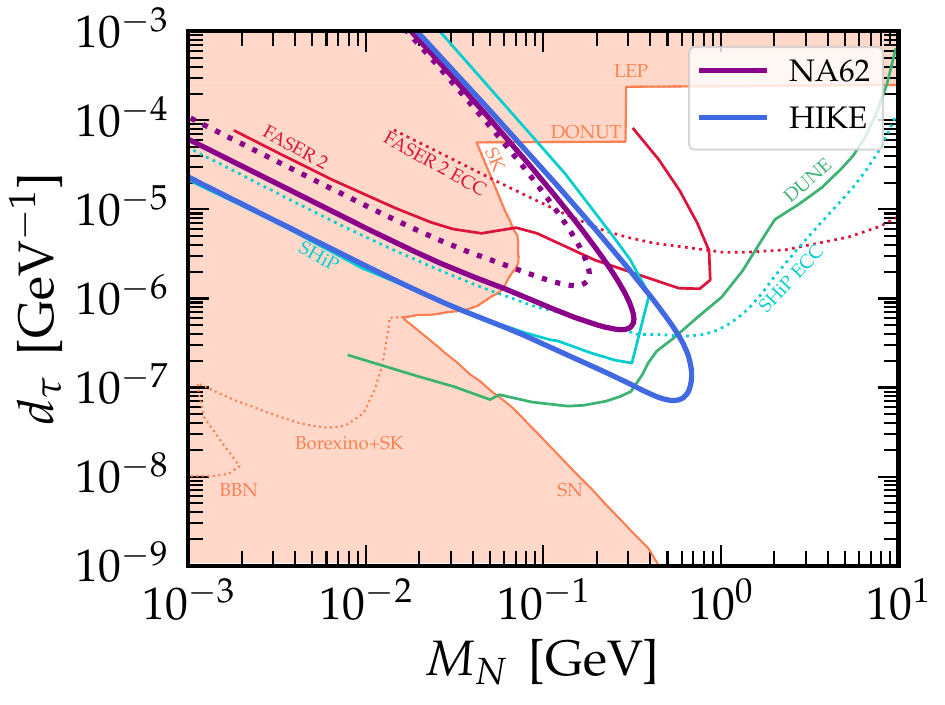}
	\caption{Isocontours for $N_{\rm evts}=3$ signal events induced by the sterile-active dipole operator of Eq.\,\eqref{eq:dipole_active-sterile} expected at NA62 and HIKE.
or NA62 solid and dotted lines are as in Fig.~\ref{fig:sensInelastic}.
 The expected sensitivity of the proposed SHiP and FASER 2 experiments are taken respectively from~\cite{Magill:2018jla} and~\cite{Jodlowski:2020vhr} considering photon signals in the decay vessel (solid) or in the  ECC detector (dotted), see main text for details.
 The expected sensitivity of the proposed DUNE experiment is taken from~\cite{Ovchynnikov:2022rqj} (combining mono-$\gamma$, double-bang, and di-lepton signatures). 
 Current constraints are represented as colored areas and are taken from LEP~\cite{Magill:2018jla}, LSND~\cite{Magill:2018jla}, NOMAD~\cite{Magill:2018jla}, MiniBooNE~\cite{Magill:2018jla}, Super-Kamiokande atmospheric neutrino data (SK)~\cite{Gustafson:2022rsz}, Borexino and Super-Kamiokande (Borexino+SK)~\cite{Plestid:2020vqf}, DONUT~\cite{DONUT:2001zvi}, CHARM-II~\cite{Coloma:2017ppo}, BBN~\cite{Brdar:2020quo} and Supernova SN1987A (assuming flavor universal couplings)~\cite{Chauhan:2024nfa}.
}
\label{fig:sensActiveSterile}
\end{figure}

\subsection{Results}

We finally present the sensitivities of the various experiments under consideration 
by imposing a minimum photon energy $E_{\gamma}>1$\,GeV and by requiring at least three signal events, $N_{\rm evts}\ge 3$. In practice, we assume that background rates can 
be reduced to a negligible level. For what concerns the case of FASER 2, we refer to\,\cite{Dienes:2023uve} for a thorough discussion of the potential background sources and the way to mitigate them.
As regarding the beam dump mode of NA62, the 
only SM particles produced from the interaction of the beam with the target material which are able of reaching the NA62 decay volume are neutrinos and muons, whose interactions could potentially generate background events for our mono$-\gamma$ analysis. In particular, for operation in beam dump mode the halo muon flux turns out to be the dominant source 
of events in the decay volume\,\cite{Rosenthal:2019qua}.
While a detailed study of this background source is beyond the scope of our exploratory study
and require a more dedicated experimental work, we observe that its rate might be kept well under control. The dipole magnets which are used in the Kaon mode for the beam modulation are in fact now used to sweep the muons away from the decay volume acceptance, see {\it e.g.}\,\cite{Jerhot:2023dhn,NA62:2023qyn},
allowing to strongly mitigate the halo muon flux.
Furthermore, there are other ways with which halo muon rates could be substantially reduced. For example, detectors preceding the decay volume that are not used at the analysis level in the beam-dump mode can be used to implement an upstream veto, see again\,\cite{Jerhot:2023dhn,NA62:2023qyn}.
Background rates will also be impacted by the chosen threshold for the energy of the photons.
The requirement  $E_\gamma>1\,$GeV has been chosen following NA62 selection in searches performed in Kaon mode of invisible $\pi^0$ decay\,\cite{NA62:2020pwi}, where the threshold requirement for detecting photons by $\pi^0$ decay, which are to be vetoed,  is set to this value.
Softer photons could in principle be identified, although a better energy resolution and a more efficient background removal is expected to be obtained with harder photons, see {\it e.g.}\,\cite{NA62:2023olg}.
In order to highlight the impact of a different choice of photon energy threshold, we also show results where we reduce or increase this value to $E_{\gamma}>0.5,\,2\,$GeV.

We present our results for the operator of Eq.\,\eqref{eq:dipole_sterile-sterile} in Fig.~\ref{fig:sensInelastic}. The colored regions correspond to the constraints discussed in Sec.~\ref{sec:Currentlimits},
while the dashed black lines correspond to the thermal relic target, see the discussion in Sec.\,\ref{sec:Currentlimits}.
We choose three values for the relative mass splitting, $\delta=10^{-1},10^{-2},5\times10^{-3},$ representative of compressed mass spectra scenarios. Overall, we find that NA62 in beam dump mode is able to probe regions of the parameter space not yet excluded by current data. 
Interestingly, this is true also for $N_{\rm POT}=10^{17}$, which roughly corresponds  to the dataset currently collected in beam dump mode.
Therefore, a dedicated search of the mono-photon signature with current data could already lead to interesting results.
Moreover, NA62 will be able to test some region of the parameter space where the lightest dark state can account for all the cosmological DM abundance via thermal freeze-out.
Note that the sensitivity of NA62 is very similar to the one of FASER 2. 
The proposed experiments HIKE and especially SHiP will significantly extend this reach\,\footnote{The sensitivities of SHiP are slightly different from the ones in~\cite{Barducci:2022gdv} mainly due to an update in the geometry of the experiment, see Sec.~\ref{sec:exp}, and to a smaller extent due to the modelling of the meson production. In particular, we have included in the analysis the pseudoscalar mesons $\pi^{0},\eta,\eta^{\prime}$, which play however a subdominant role, as already estimated in~\cite{Barducci:2022gdv}. For FASER\,2 the main difference with~\cite{Barducci:2022gdv} is the choice of the energy threshold. In~\cite{Barducci:2022gdv}, inspired by other FASER\,2 analyses, we explored more conservative thresholds, from 10 GeV up to 200 GeV. Moreover, we had conservatively considered $N_2$ decays only in the decay volume. Adding also the tracking system, as suggested in~\cite{Dienes:2023uve}, raises $L_f$ from $L_f=630$ m to $L_f=640$ m.}. 
For these compressed spectra, the energy threshold plays an important role. In fact, for small mass splittings the typical energy of the photon is $E_{\gamma}\simeq 2\,\delta\,P_{\chi_2}$, with $P_{\chi_2}$ being the momentum of $\chi_2$ in the laboratory frame\,\cite{Barducci:2022gdv}. 
Clearly, for small values of $\delta$ an increasing larger fraction of the events falls below the energy threshold.

In Fig.~\ref{fig:sensInelasticEcut} we then show our results obtained by varying the photon energy threshold by a factor of two around the reference value, {\it i.e.} $E_{\gamma}\ge 0.5$ GeV and $E_{\gamma}\ge 2$ GeV. As evident, this has a strong impact on the sensitivities for $\delta\lesssim5\times 10^{-3},$
 except for FASER 2, which sensitivity remains almost unchanged for energy thresholds in the $\mathcal{O}$(GeV) range due to the large center-of-mass energy.
We stress that the variation of the energy threshold that we have explored is only indicative, and its purpose is to remark the important role of this ingredient. In fact, our results motivate detailed studies on the mono-$\gamma$ signature from the experimental collaborations, in order to  quantify the optimal energy threshold (as well as general selections) needed to maximize the signal keeping at the same time negligible background rates.

In Fig.~\ref{fig:sensActiveSterile} we present our results for the active-sterile neutrino dipole of Eq.\,\eqref{eq:dipole_active-sterile}. We consider separately the cases of $e,\mu,\tau$ flavors. 
For SHiP and FASER\,2 the relevant sensitivities have already been derived in~\cite{Magill:2018jla} and~\cite{Jodlowski:2020vhr}, which we directly present\footnote{In addition to a decay volume, both the SHiP and FASER\,2 experiments are also instrumented with an emulsion cloud chamber (ECC) detector, dedicated to study neutrinos. The sterile neutrino $N_1$ can be produced via dipole interactions by upscattering of active neutrinos, generated by proton collisions, in the ECC or around the decay volume. This production mechanism adds to the one provided by meson decays. In addition, both the ECC detector and the decay volume can be used to detect photons from the decay of $N_1$. Separated sensitivities for these two searches are shown in  Fig.~\ref{fig:sensActiveSterile}.} in Fig.~\ref{fig:sensActiveSterile}.
 While existing bounds depend significantly on the flavor structure, the sensitivity of NA62 is almost independent of it. We find that NA62 and a possible upgrade as the proposed HIKE experiment offer promising opportunities to detect the sterile-active dipole interaction, by extending beyond what is tested by current available data.

\section{Conclusions}
\label{sec:conc}

Dipole interactions are well-motivated and among the most studied portals
connecting the Standard Model and dark sectors, possibly associated with the origin of Dark Matter and neutrino masses. By considering 
both a dipole interaction between two beyond the Standard Model states as well as a dipole interaction connecting a dark state with active neutrinos, we have extended the results  of\,\cite{Barducci:2022gdv} by studying the reach of the NA62 experiment running in beam dump mode. In particular, we propose 
to search for mono-$\gamma$ final states arising from
dark sector particles produced by
the exotic decay of Standard Model mesons.
We have demonstrated that, 
even with the already collected dataset, NA62  has a significant potential to discover dipole interactions, probing regions of parameter space not yet excluded by other experiments. Searches performed with the planned integrated dataset will further extend this reach.
We have presented the relevant sensitivity reach in the parameter space of these models, and compared with the ones of a possible extension of NA62 with a larger luminosity, and the proposed experiments FASER 2 and SHiP.

\section*{Acknowledgements}
We thank Marco Sozzi for useful discussions regarding NA62. We are grateful to Ryan Plestid for clarifications on his previous work~\cite{Magill:2018jla} and to Maksym Ovchynnikov for help with the {\tt SensCalc} code~\cite{Ovchynnikov:2023cry}. The work of EB is partly supported by the Italian INFN program on Theoretical Astroparticle
Physics (TAsP), by ``Funda\c{c}\~ao de Amparo \`a Pesquisa do Estado de S\~ao Paulo”
(FAPESP) under contract 2019/04837-9, as well as by Brazilian “Conselho
Nacional de Deselvolvimento Cient\'ifico e Tecnol\'ogico” (CNPq).
The work of CT is supported 
by the Italian Ministry of University and Research (MUR) 
via the PRIN 2022 project n.~2022K4B58X -- AxionOrigins.
MT acknowledges support from the project ``Theoretical Astroparticle Physics (TAsP)'' funded by the INFN and from the research grant ``Addressing systematic uncertainties in searches for dark matter No. 2022F2843L'' funded by MIUR.


\appendix
\section{Decay widths}\label{app:decay_widths}
We present in this appendix some useful equations regarding the decay widths used in our computations. 

\subsection{Heavy dark state decay mediated by the dipole}
The dipole operators of Eq.~\eqref{eq:dipole_sterile-sterile} and Eq.\,\eqref{eq:dipole_active-sterile} generate the decay of the heavier state into the lighter state and one photon. We obtain:
\be
\begin{split}
    \Gamma(\chi_2 \to \chi_1 \gamma) & =d^2\,  \frac{ M_1^3}{2\pi} \frac{\delta^3 ( 2+\delta)^3}{(1+\delta)^3} ,\\
    \Gamma(N_1 \to \nu_i \gamma) & = d_i^2\, \frac{M_N^3}{4\pi} ,
\end{split}
\ee
where in the first line $M_1$ is the mass of the lighter dark state and $\delta = (M_2 - M_1)/M_1$, with $M_2$ the mass of the heavier dark state. In the second line, on the other hand, $M_{N}$ is the mass of the Dirac fermion $N_1$.

For the operator in Eq.\,\eqref{eq:dipole_active-sterile}, $N_1$ and its antiparticle are different objects, so we must also consider the decay $\bar{N}_1 \to \bar{\nu}_i \gamma$ in our computations.

\subsection{Vector meson decay} 
The decay width of a vector meson $V= \left\{\rho, \omega, J/\psi, \Upsilon, \phi\right\}$ into $\chi_1\chi_2.$ reads:
\begin{align}
\label{eq:Vtochi1chi2}
\Gamma(V\rightarrow\chi_1\chi_2) &= \frac{d^2 e^2 Q_q^2 f_V^2 M_V}{6\pi} \left(1-\frac{(M_2 - M_1)^2}{M_V^2}\right)^{3/2}\nonumber\\
&\times\left( 1+\frac{2(M_2 + M_1)^2}{M_V^2} \right) \left( 1-\frac{(M_2 + M_1)^2}{M_V^2}\right)^{1/2}\,,
\end{align}
where $e$ is the electron charge, $Q_q$ the electric charge (in units of $e$) of the quark $q$ composing the meson $V\sim q\bar{q},$ $M_V$ the mass of the meson and $f_V$ its decay constant (reported \emph{e.g.} in~\cite{Bertuzzo:2020rzo,Barducci:2023hzo}).  
For the operator in Eq.\,\eqref{eq:dipole_active-sterile}, the decay width for $V \to N_1 \nu_i$ is obtained, in the limit of massless active neutrino, by simply setting $M_1\rightarrow 0$, $M_2\rightarrow M_N$ and $d\rightarrow d_i/\sqrt{2}$ in Eq.\,\eqref{eq:Vtochi1chi2}. Also in this case, we must consider the production of $N_1$'s antiparticle via $V \to \bar{N}_1 \nu_i$, which has the same width as $V \to N_1 \nu_i$.

\subsection{Pseudoscalar meson decay}

We next discuss the decay of a pseudoscalar meson $\mathcal{M}_P^0= \left\{\pi^0, \eta, \eta^{\prime}\right\}$ via the anomaly with subsequent photon splitting.
The kinematic is
\be
P^0 \to \gamma(p_1) \Psi_{M_1}(p_2) \Psi_{M_2}(p_3) \ .
\ee

The squared matrix element summed over the final polarizations is:
\begin{multline}
|{\cal \overline M}|^2 = \frac{d^2 e^4}{4\pi^4 F_P^2 m_{23}^4} \left[ M_P^4\left( m_{23}^2(M_2^2+2M_1M_2-M_1^2)-(M_1^2-M_2^2)^2 \right) - \right. \\ \left.
m_{23}^4(M_1^4-2M_1^2 m_{12}^2- m_{23}^2 (M_1+M_2)^2 +2 m_{12}^4 -2 m_{12}^2M_2^2 +2 m_{12}^2 m_{23}^2 +M_2^4 ) - \right. \\ 
\left. 2m_{23}^2 M_P^2 \left( (M_1^2-M_2^2)(M_2^2-m_{12}^2) +M_2m_{23}^2(2M_1+M_2) -m_{12}^2m_{23}^2\right)
\right],
\end{multline}
with $p_{ij}=p_i+p_j$ and $m_{ij}^2=p_{ij}^2$,
 $M_P$ the mass of the pseudoscalar meson and $F_P$ the  decay constant ($F_{\pi}\simeq 92\,{\rm MeV}$ and for $\eta,\eta^{\prime}$ see~\cite{Bali:2021qem}).
 The total width is
 \be
 \Gamma = \frac{1}{(2\pi)^3\,32M_P^3}\int dm_{12}^2 dm_{23}^2 |{\cal \overline M}|^2.
\ee
Also in this case, when considering the operator in Eq.\,\eqref{eq:dipole_active-sterile}, the width is obtained with the straightforward changes explained above. Once more, also production of $\bar{N}_1$ must be considered.

\subsection{Charged pseudoscalar meson decay}

The decay of a charged pseudoscalar meson is relevant in the case of an active/sterile dipole where the $P^+$ meson decays leptonically and the dipole triggers the final $\nu$ splitting into a sterile  neutrino and photon pair.
We consider $P^+ = \left\{B^+, D^+, D_s, K^{\pm}\right\}.$

The kinematic is
\be
P^+ \to \gamma(p_1) \ell^+(p_2) \psi_{M}(p_3)\,.
\ee

The matrix element squared is:
\begin{multline}
|{\cal \overline M}|^2 =\frac{16 d_i^2 G_F^2 f_P^2 |V_{CKM}|^2}{(M_P^2+M_l^2+M_N^2-m_{23}^2-m_{12}^2)^2} \left(
M_P^2 + M_l^2-m_{12}^2 - m_{23}^2  \right)
\left[
m_{12}^6-m_{12}^4(2M_P^2+2M_N^2 +M_l^2-
\right. \\
\left.
2m_{23}^2)+
m_{12}^2\left(M_N^4+2M_N^2(M_P^2+M_l^2-m_{23}^2) +m_{23}^4-m_{23}^2(2M_P^2+M_l^2)+M_P^4
\right)
\right.
\\
\left.
+M_l^2\left(
-M_N^4-M_P^2(M_N^2-M_l^2+m_{23}^2)+M_N^2m_{23}^2+M_P^4
\right)
\right],
\end{multline}
 where $M_l$ is the mass of the lepton, $G_F$ is the Fermi constant, $M_P$ is the mass of the meson, $F_P$ is the  decay constant~\cite{FlavourLatticeAveragingGroupFLAG:2021npn} and $V_{CKM}$ is the CKM matrix element involved in the relevant transition.

\newpage

\providecommand{\href}[2]{#2}\begingroup\raggedright\endgroup

\end{document}